\documentclass[11pt]{article}
\usepackage{moriond}
\usepackage{enumerate}

\def\be{\begin{equation}}
\def\ee{\end{equation}}
\def\bea{\begin{eqnarray}}
\def\eea{\end{eqnarray}}

\def\ie{{\it i.e.}}
\def\eg{{\it e.g.}}

\newcommand{\Photo}{}

\begin{document}
\vspace*{4cm}
\title{HIGGS COUPLINGS AFTER MORIOND}

\author{B\'ERANGER DUMONT}

\address{Laboratoire de Physique Subatomique et de Cosmologie, UJF Grenoble 1,
CNRS/IN2P3, INPG, \\ 53 Avenue des Martyrs, F-38026 Grenoble, France}

\maketitle\abstracts{
Performing fits to all publicly available data, we analyze the extent to which the latest results from the LHC and Tevatron (including new results presented at the Rencontres de Moriond) constrain the couplings of the Higgs boson-like state at $\sim 125$~GeV, as well as possible decays into invisible particles.
%To this end we assume that tree-level Higgs couplings to the up-quarks, down-quarks and vector bosons, relative to the SM are free parameters. We also explore the possibility that the net Higgs to $\gamma\gamma$ and $gg$ couplings have extra loop contributions coming from Beyond-the-Standard Model physics. Finally, we study cases in which the Higgs decays into invisible or undetected particles. We find that the SM provides a very good fit to the data ($p$-value $\sim 0.65$) and that there is no preference for New Physics contributions to the effective couplings of the Higgs.
}

\section{Setup of the analysis}

The recent discovery~\cite{atlas:2012gk,cms:2012gu} of a new particle with mass around 125 GeV and properties consistent with a Standard Model (SM) Higgs boson is clearly the most significant news from the Large Hadron Collider (LHC). This discovery was supported by consistent measurements by the CDF and D0 collaborations at the Tevatron~\cite{Aaltonen:2013kxa} and completes our picture of the SM.
However, the SM leaves many fundamental questions open---perhaps the most pressing issue being that the SM does not explain the value of the electroweak scale, \ie\  the Higgs mass, itself. Clearly, a prime goal after the discovery is to thoroughly test the SM nature for this Higgs-like signal.

The experimental collaborations have presented detailed results for several different channels, including $\gamma \gamma$, $ZZ^{(*)}$, $WW^{(*)}$, $b\bar{b}$, $\tau \tau$ final states and invisible decays. With these measurements, a comprehensive study of the properties of the Higgs-like state becomes possible and has the potential for revealing whether or not the Higgs sector is as simple as envisioned in the SM, see \eg~\cite{Belanger:2012gc} and references therein. Here, we provide an update of the Higgs couplings fits of~\cite{Belanger:2012gc}, based on the most recent results from the LHC and Tevatron presented at this conference. (A long paper~\cite{Belanger:inprep} is in preparation.)

Following~\cite{Belanger:2012gc}, we parametrize deviations from the SM couplings by the Lagrangian
\be
   {\cal L} =  g\left[  C_V \left(m_W W_\mu W^\mu+{m_Z\over \cos\theta_W} Z_\mu Z^\mu \right)  
   - C_U {m_t\over 2m_W} \bar tt  - C_D    {m_b\over 2m_W} \bar bb - C_D {m_\tau\over 2 m_W}\bar\tau\tau \right ]H\,. 
   \label{ourldef}
\ee
where we introduce scaling factors $C_I$. We assume a single $C_W=C_Z \equiv C_V$ and that the reduced couplings to up-type and down-type fermions, $C_U$ and $C_D$, are independent parameters.
In general, the $C_I$ can take on negative as well as positive values; there is one overall sign ambiguity which we fix by taking $C_V>0$.

In addition to the tree-level couplings given above, the $H$ has couplings to $\gamma\gamma$ and $gg$ that are first induced at one loop and are completely computable in terms of $C_U$, $C_D$ and $C_V$ if only loops containing SM particles are present. We define $\overline C_\gamma$ and $\overline C_g$ to be the ratio of these couplings so computed to the SM (\ie\ $C_U=C_V=C_V=1$) values.
However, in some of our fits we will also allow for additional loop contributions $\Delta C_\gamma$ and $\Delta C_g$ from new particles; in this case $C_\gamma = \overline C_\gamma + \Delta C_\gamma$ and $C_g = \overline C_g + \Delta C_g$. The largest set of independent parameters in our fits is thus $C_U,~C_D,~C_V,~\Delta C_\gamma,~\Delta C_g$.

The experimental results are given in terms of signal strengths $\mu(X,Y)$, the ratio of the observed rate for some process $X\to H \to Y$ relative to the prediction for the SM Higgs. As in~\cite{Belanger:2012gc}, we adopt the simple technique of computing the $\chi^2$ associated with a given choice of the input parameters following the standard definition: 
\be
  \chi^2=\sum_k {(\overline\mu_k-\mu_k)^2\over \Delta \mu_k^2}\,,
\label{chisqdef}
\ee
where $k$ runs over all the experimentally defined production/decay channels employed, $\mu_k$ is the observed signal strength for a channel $k$, $\overline\mu_k$ is the value predicted for that channel for a given choice of parameters and $\Delta \mu_k$ is the experimental error for that channel.  The $\overline\mu_k$ associated with each experimentally defined channel is further decomposed as $\overline\mu_k=\sum T^i_k \,\widehat \mu_i$, where the $T^i_k$ give the amount of contribution to the experimental channel $k$ coming from the theoretically defined channel $i$ and $\widehat \mu_i$ is the prediction for that channel for a given choice of $C_U$, $C_D$, $C_V$ and (for fits where treated as independent) $C_\gamma$ and $C_g$.  
For the computation of the $\widehat\mu_i$ including NLO corrections we follow the procedure recommended by the  LHC Higgs Cross Section Working Group~\cite{LHCHiggsCrossSectionWorkingGroup:2012nn}.
\section{Results}

In order to probe the SM nature of the observed Higgs boson, we perform the following fits:
\begin{enumerate}[I)]
\itemsep0em
\item fit of $\Delta C_\gamma$ and $\Delta C_g$, assuming $C_U=C_D=C_V=1$;
\item fit of $C_U$, $C_D$ and $C_V$, assuming $\Delta C_\gamma = \Delta C_g = 0$;
\item fit of $C_U$, $C_D$, $C_V$, $\Delta C_\gamma$ and $\Delta C_g$, restricted to $C_U>0$ and $C_D>0$.
\end{enumerate}

\subsection{Fit I: $\Delta C_\gamma$ and $\Delta C_g$}
In this first case, we allow for additional new physics contributions  to the $\gamma\gamma$ and $gg$ couplings, parameterized by $\Delta C_\gamma$ and $\Delta C_g$, coming from loops involving non-SM particles. 
This fit, which we refer to as Fit I, is designed to determine if the case where all tree-level Higgs couplings 
are equal to their SM values can be consistent with the data. For example, such a fit is relevant in the context of UED models where the tree-level couplings of the Higgs are SM-like~\cite{Petriello:2002uu,Belanger:2012mc}.

\begin{figure}[t]\centering
\includegraphics[scale=0.4]{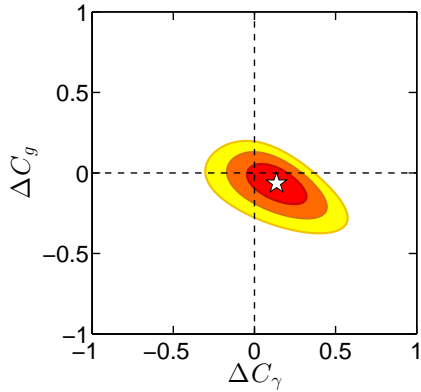}
\caption{Two parameter fit of $\Delta C_\gamma$ and $\Delta C_g$, assuming $C_U=C_D=C_V=1$ (Fit I). 
The red, orange and yellow ellipses show the 68\%, 95\% and 99.7\%  CL regions, respectively. 
The white star marks the best-fit point.
\label{CPadd-CGadd} }
\end{figure}

Figure~\ref{CPadd-CGadd} displays the results of this fit in the  $\Delta C_g$ versus $\Delta C_\gamma$ plane.  
The best fit is obtained for $\Delta C_\gamma = 0.14$, $\Delta C_g = -0.06$,  and has $\chi^2_{\rm min}=17.6$ for 20 degrees of freedom (d.o.f.), giving a $p$-value of $0.61$. We note that the SM (\ie\ $C_U=C_D=C_V=1$, $\Delta C_\gamma=\Delta C_g=0$) has $\chi^2= 19.0$ for 22 d.o.f., implying a $p$-value of $0.65$.

\subsection{Fit II: $C_U$, $C_D$ and $C_V$}
Next, we let  $C_U$, $C_D$ and $C_V$ vary, assuming there are no new particles contributing to the effective 
Higgs couplings to gluons and photons, \ie\  we take $\Delta C_\gamma=\Delta C_g=0$, implying $C_\gamma$ and $C_g$ as computed from the SM particle loops. Such parametrization is relevant in the context of 2HDM with a heavy charged Higgs, that does not contribute to the loop-induced $H \rightarrow \gamma\gamma$ process.

\begin{figure}[t]\centering
\includegraphics[scale=0.4]{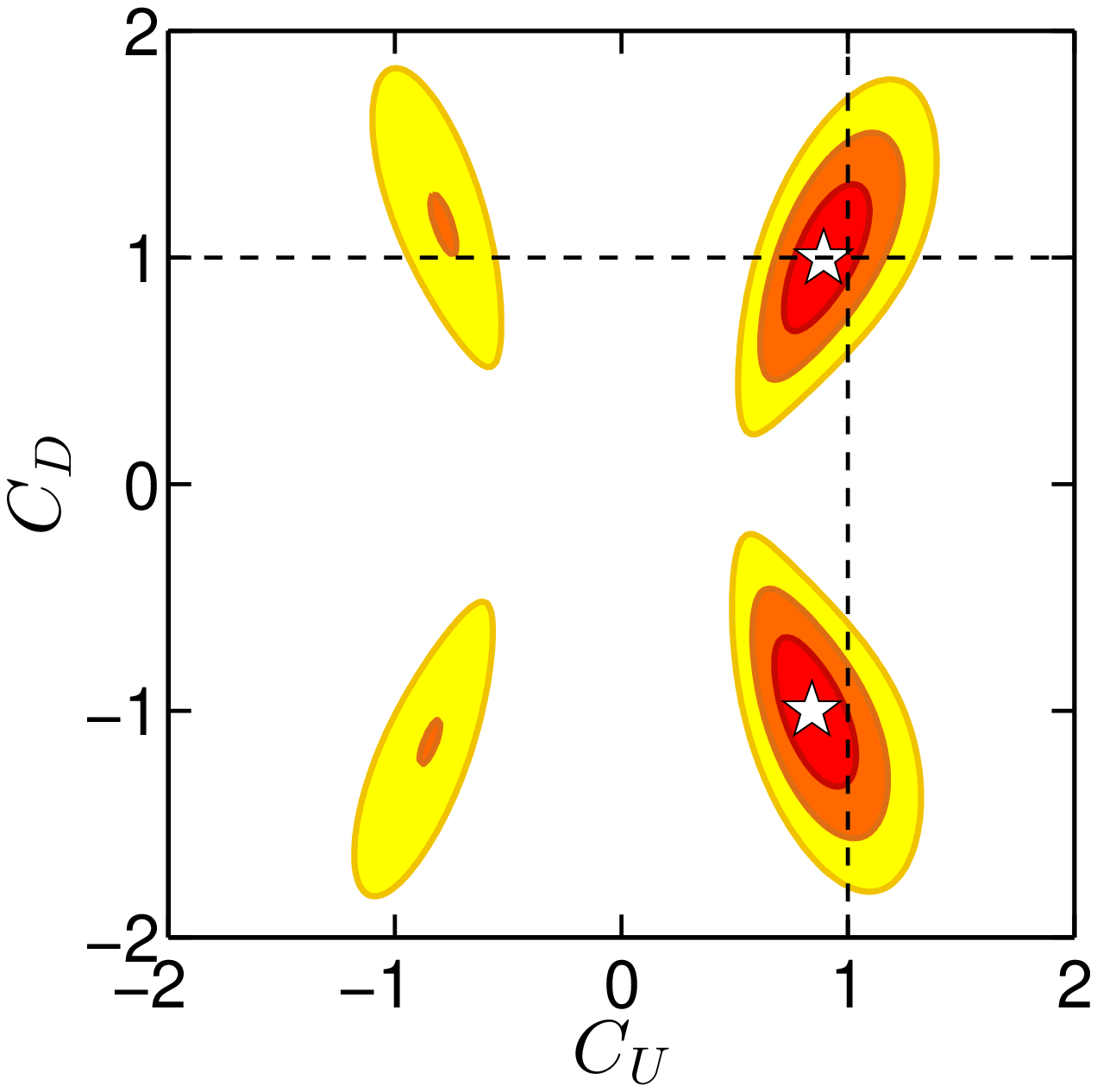}
\includegraphics[scale=0.4]{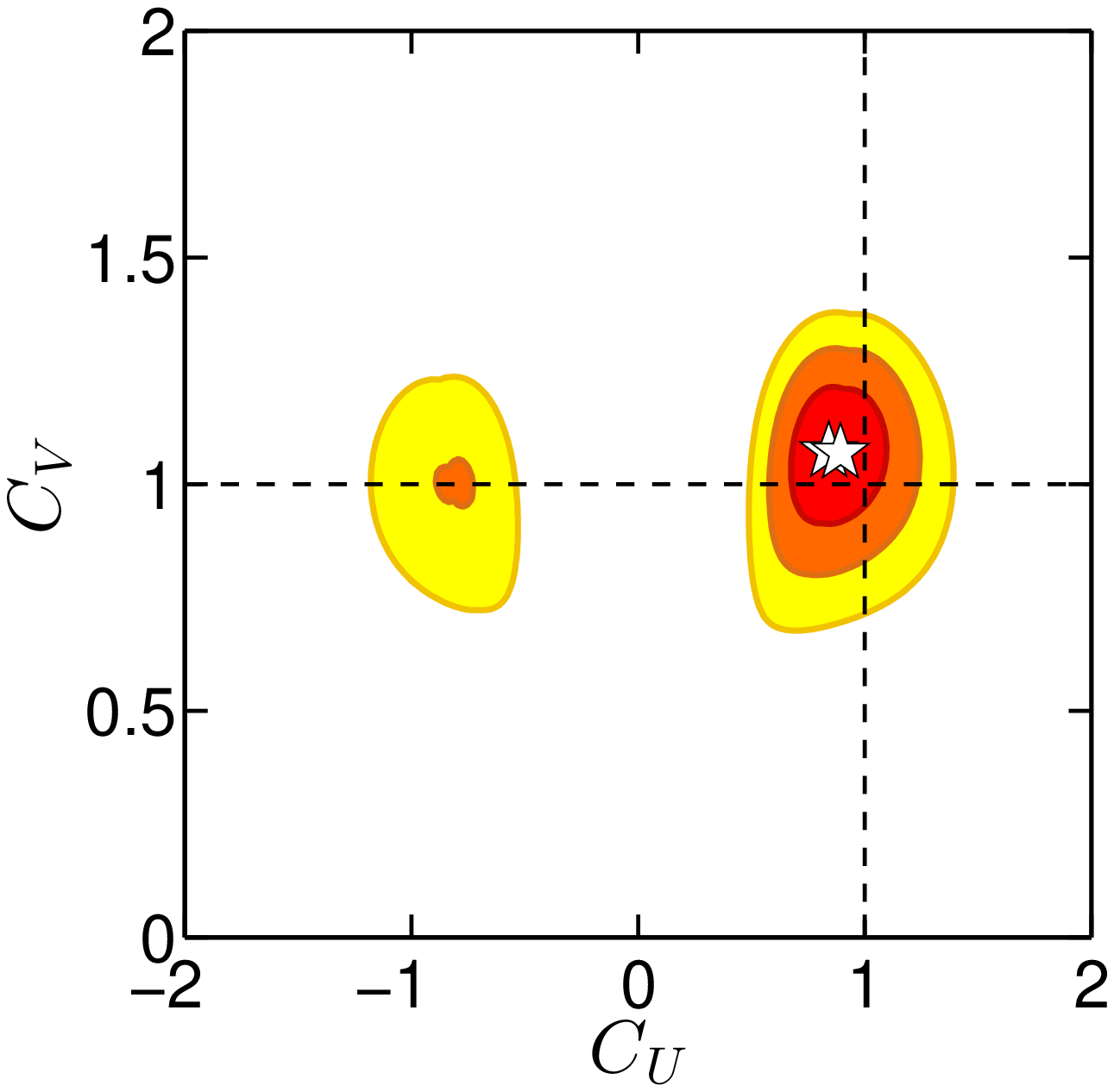}
\caption{Two-dimensional $\chi^2$ distributions for the three parameter fit of $C_U$, $C_D$, $C_V$ with $C_\gamma$ and $C_g$ as computed in terms of $C_U, C_D, C_V$ (Fit II). Color code as in the previous figure.
\label{CU-CD-CV} }
\end{figure}

Our results are shown in Fig.~\ref{CU-CD-CV}. We consider two best fits points, having positive and negative $C_D$. The one with $C_D>0$ is located at $C_U = 0.89$, $C_D = 0.99$, $C_V = 1.07$, and $C_U = 0.84$, and has $\chi^2_{\rm min} = 17.7$ for 19 d.o.f., giving a $p$-value of $0.54$. (The one with $C_D<0$ is almost equivalent---the sign of $C_D$ only affects mildly the loop-induced processes---and has $\chi^2_{\rm min} = 17.6$.)
Contrary to the situation at the end of 2012~\cite{Belanger:2012gc} we note that the regions having $C_U<0$, in which the $H \rightarrow \gamma\gamma$ rate is significantly enhanced, are disfavored at the level of $2.4\sigma$. This mainly comes from the update of the CMS $H \rightarrow \gamma\gamma$ results presented at this conference~\cite{CMS-PAS-HIG-13-001}.

\subsection{Fit III: $C_U$, $C_D$, $C_V$, $\Delta C_\gamma$ and $\Delta C_g$}
\begin{figure}[t]\centering
\includegraphics[scale=0.4]{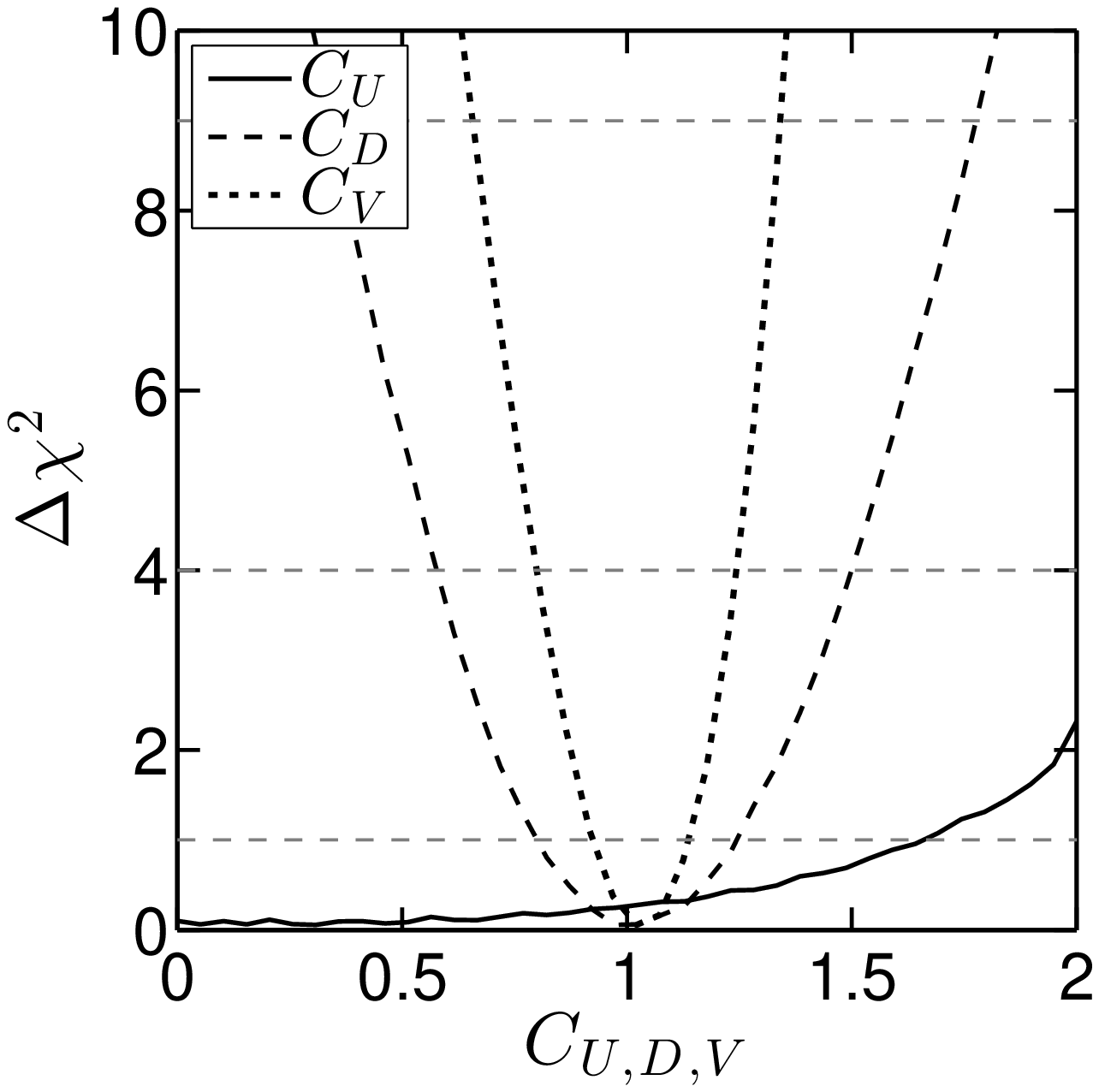}
\includegraphics[scale=0.4]{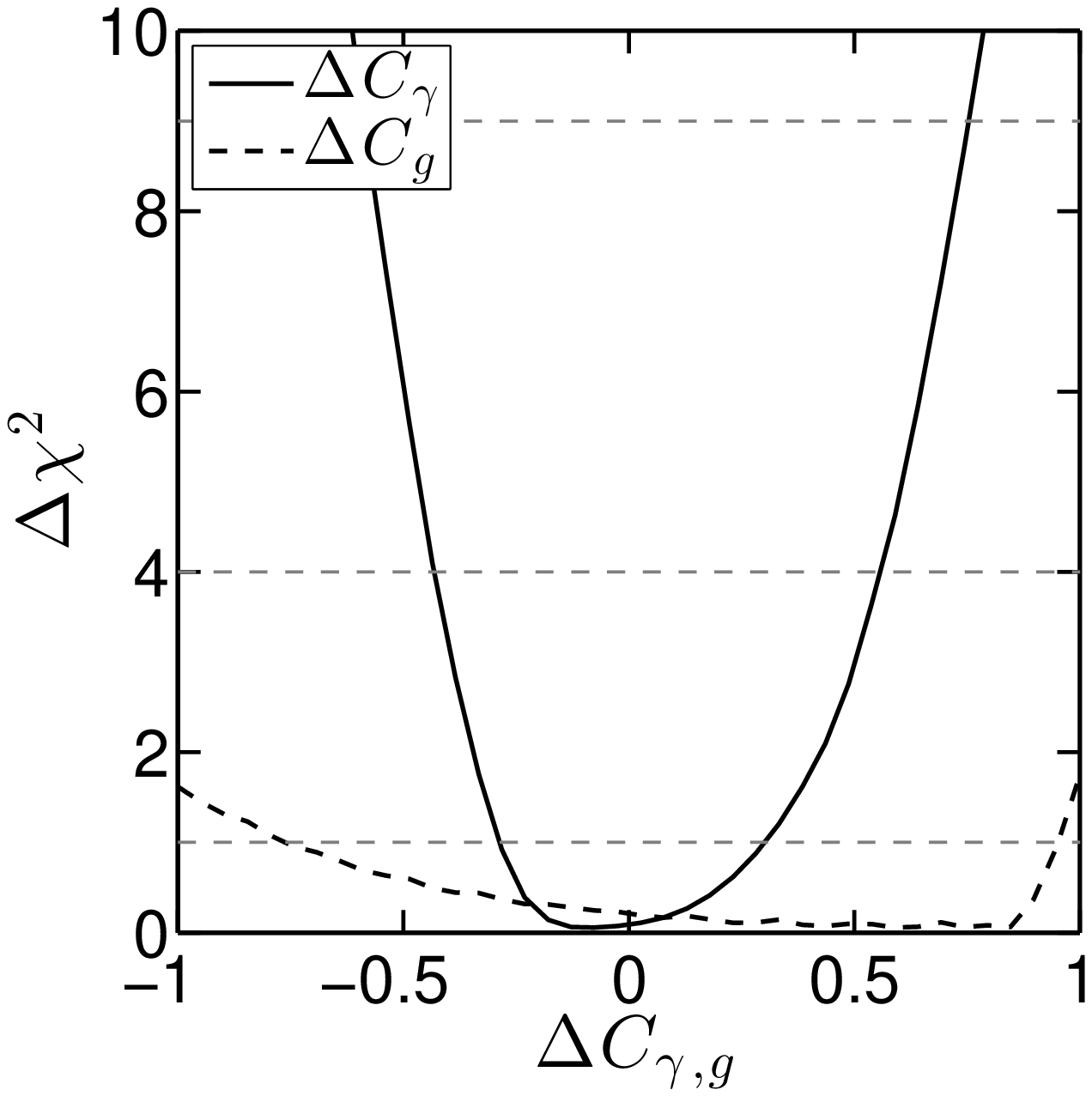}
\caption{One-dimensional $\chi^2$ distributions for the five parameter fit of $C_U$, $C_D$, $C_V$, $\Delta C_\gamma$ and $\Delta C_g$ (Fit III).
\label{CU-CD-CV-CPadd-CGadd} }
\end{figure}

Finally, in Fit III, we allow for new particles entering the loop, parametrized by $\Delta C_\gamma$ and $\Delta C_g$, in addition to the tree-level parameters $C_U$, $C_D$ and $C_V$, leading therefore to five free parameters. This encompasses a very broad class of models.
The associated 1d plots are given in Fig.~\ref{CU-CD-CV-CPadd-CGadd}.

The main differences as compared to Fit~{\bf II} is that $C_U$ is only weakly constrained by the data. Allowing for  additional contributions to $H \rightarrow \gamma\gamma$ and gluon fusion, parametrized by $\Delta C_\gamma$ and $\Delta C_g$, account for the observed rates of these processes. $C_U$ is then only determined from the results on associated top pair production.
The best fit is at $C_U = 0$, $C_D = 1.02$, $C_V = 1.04$, $\Delta C_\gamma = -0.16$, $\Delta C_g = 0.82$ and has $\chi^2_{\rm min} = 17.2$ for 17 d.o.f., giving a $p$-value of $0.44$.

\subsection{Limits on invisible decays of the Higgs}
\begin{figure}[t]\centering
\includegraphics[scale=0.4]{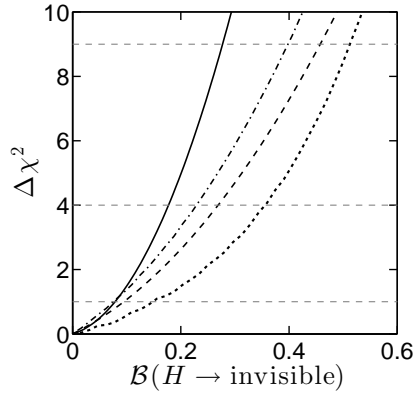}
\caption{$\Delta\chi^2$ distributions for the branching ratio of invisible Higgs decays. The full, dashed, and dotted lines correspond, respectively, to the cases of  1) SM couplings, 2) arbitrary $\Delta C_\gamma$ and $\Delta C_g$ (Fit I), and 3)~deviations of $C_U, C_D, C_V$ from 1 (Fit II). We also show as dash-dotted line the variant of case 3) with $C_U, C_D>0$ and $C_V \le 1$.
\label{1d-fit-invisible} }
\end{figure}

The current measurements can also be used to derive limits on the invisible decays of the Higgs, as was done in~\cite{Belanger:2013kya}. We also include in our fit the ATLAS limit on $ZH \rightarrow \ell \ell + {\rm invisible}$~\cite{ATLAS-CONF-2013-011}. Our results are shown in Fig.~\ref{1d-fit-invisible}. The 95\% CL limits on ${\cal B}(H \rightarrow {\rm invisible})$ range from 18\% to 35\%, depending on our assumption on the couplings of the Higgs boson. 

\section*{Acknowledgments}
I would like to sincerely thank Genevi\`eve B\'elanger, Ulrich Ellwanger, John F. Gunion and Sabine Kraml for the rewarding collaboration on~\cite{Belanger:2012gc,Belanger:inprep,Belanger:2013kya} and the opportunity to present these studies on behalf of the group at the Rencontres de Moriond.
Partial financial support by IN2P3 under contract PICS FRÐUSA No. 5872 and by US DOE grant DE-FG03-91ER40674 is gratefully acknowledged.

\end{document}